\documentclass[12pt]{article}
\usepackage{a4}
\usepackage{amsfonts}
\usepackage{amsmath,amssymb}

\usepackage{graphicx}

\def\un#1{\relax\ifmmode\@@underline#1\else
        $\@@underline{\hbox{#1}}$\relax\fi}

\let\du=\du                     

\def\f{\phi}

\def\l{\lambda}
\def\m{\mu}
\def\n{\nu}

\def\s{\sigma}

\def\F{\Phi}

\def\car{{\cal R}}

\def\cy{{\cal Y}}

\def\bo{{\raise-.3ex\hbox{\large$\Box$}}}               
\def\pa{\partial}                                       
\def\TH{{\raise.2ex\hbox{$\displaystyle \bigodot$}\mskip-4.7mu \llap H \;}}
\def\face{{\raise.2ex\hbox{$\displaystyle \bigodot$}\mskip-2.2mu \llap {$\ddot
        \smile$}}}                                      

\def\abs#1{\left| #1\right|}                    
\def\leftrightarrowfill{$\mathsurround=0pt \mathord\leftarrow \mkern-6mu
        \cleaders\hbox{$\mkern-2mu \mathord- \mkern-2mu$}\hfill
        \mkern-6mu \mathord\rightarrow$}
\def\dvec#1{\vbox{\ialign{##\crcr
        \leftrightarrowfill\crcr\noalign{\kern-1pt\nointerlineskip}
        $\hfil\displaystyle{#1}\hfil$\crcr}}}           

\def\frac#1#2{{\textstyle{#1\over\vphantom2\smash{\raise.20ex
        \hbox{$\scriptstyle{#2}$}}}}}                   
\def\sfrac#1#2{{\vphantom1\smash{\lower.5ex\hbox{\small$#1$}}\over
        \vphantom1\smash{\raise.4ex\hbox{\small$#2$}}}} 
\def\bfrac#1#2{{\vphantom1\smash{\lower.5ex\hbox{$#1$}}\over
        \vphantom1\smash{\raise.3ex\hbox{$#2$}}}}       
\def\afrac#1#2{{\vphantom1\smash{\lower.5ex\hbox{$#1$}}\over#2}}    

\def\[{\lfloor{\hskip 0.35pt}\!\!\!\lceil}
\def\]{\rfloor{\hskip 0.35pt}\!\!\!\rceil}

\def\du#1#2{_{#1}{}^{#2}}

\def\ha{{\fracmm12}}

\def\un{\underline}
\def\fracmm#1#2{{{#1}\over{#2}}}

\def\low#1{{\raise -3pt\hbox{${\hskip 0.75pt}\!_{#1}$}}}

\newskip\humongous \humongous=0pt plus 1000pt minus 1000pt

\newif\ifdtup

\newcommand{\be}{\begin{equation}}
\newcommand{\ee}{\end{equation}}
\newcommand{\nbe}{\begin{equation*}}
\newcommand{\nee}{\end{equation*}}

\newcommand{\lb}{\label}

\begin{document}

\thispagestyle{empty}

{\hbox to\hsize{
\vbox{\noindent IPMU12-0108 \hfill  September 2012 }}}

\noindent
\vskip2.0cm
\begin{center}

{\large\bf  DARK ENERGY IN MODIFIED SUPERGRAVITY}
\vglue.3in

   Sergei V. Ketov~${}^{a,b}$ and Natsuki Watanabe~${}^a$
\vglue.1in

${}^a$ {\it Department of Physics, Graduate School of Science, 
   Tokyo Metropolitan University, Hachioji-shi, Tokyo 192-0397, Japan}\\
${}^b$ {\it Kavli Institute for Physics and Mathematics of Universe, The 
           University of Tokyo, Kashiwa-shi, Chiba 277-8568, Japan}\\
\vglue.1in
ketov@phys.se.tmu.ac.jp, watanabe-natsuki1@ed.tmu.ac.jp
\end{center}

\vglue.3in

\begin{center}
{\Large\bf Abstract}
\end{center}
\vglue.1in

\noindent We propose a supersymmetric extension of the dynamical dark energy function
and the scalar (super)potential in $F(\car)$ supergravity. Our model is viable in the 
Einstein approximation, and also has an analytic (regular) scalar potential. The hidden 
sector responsible for spontaneous supersymmetry breaking is given too. 

\newpage

\section{Introduction}

The Standard ($\Lambda$CDM) Model of cosmology provides the simplest 
description of the current cosmic acceleration in agreement with all known 
observational data \cite{cmc}. However, it does not explain its origin and its 
value. The next simple models are given by {\it dynamical} Dark Energy (DE), 
such as quintessence and $f(R)$ gravity. The $f(R)$ gravity is known as the 
non-trivial, viable and consistent alternative to a cosmological constant, while
 it is classically equivalent to the quintessence \cite{stu}. An embedding (or 
derivation) of a viable $f(R)$ gravity from a more fundamental theory of gravity 
is unknown, whereas its consistency with particle physics can only be established 
in the context of a unified theory beyond the Standard Model of elementary particles.
The leading proposals for such unified theory are supersymmetry, supergravity 
and superstrings.

The supersymmetric extension of $f(R)$ gravity in curved superspace was 
proposed in ref.~\cite{Gates}, where it was dubbed $F(\car)$ supergravity. Its
structure was studied in 
refs.~\cite{our5,our2,Ketov10,our3,Watanabe,Kreview,Ketov12}, 
whereas its consistency and viability for inflation and reheating in the early 
universe was established in refs.~\cite{Ketov11,shinji}. The first application 
of $F(\car)$ supergravity to the current DE was given in ref.~\cite{kw2}.
It is worth mentioning that the cosmological $\Lambda$CDM Model {\it cannot} be
 naively extended to supergravity, since pure (Einstein) supergravity can only 
have a negative or vanishing cosmological constant.

A viable description of the current dark energy in $f(R)$ gravity imposes
certain constraints on the function $f(R)$ (see ref.~\cite{Kreview} for our
notation):
\be  \abs{f(R)-\left(-\ha R\right)} \ll \abs{R}~,\quad  
\abs{f'(R)-\left(-\ha\right)} \ll 1\quad{\rm and}\quad
 \abs{R}f''(R) \ll 1~,\lb{frc}
\ee
where the primes denote the derivatives with respect to the argument $R$ 
(the scalar curvature of spacetime), as well as the stability conditions 
\be \lb{stabi}
f'(R)<0 \quad {\rm and}\quad f''(R) >0
\ee
In particular, it means that all those models must be close to the Standard 
$\Lambda$CDM Model, while some fine-tuning is required to get the observed 
value of the present cosmic acceleration. Still, there is considerable 
(functional) freedom in the choice of the function $f(R)$ satisfying all the
criteria, see e.g., refs.~\cite{ab,hs,star3} for some explicit viable examples.
In this paper we impose more theoretical constraints on the functions $F(\car)$  
and $f (R )$ via the corresponding scalar potential in the low space-time curvature 
regime relevant to the Einstein approximation. 

Our paper is organised as follows. In Sec.~2 we briefly review the classical
correspondence between $f(R)$ gravity and quintessence \cite{eq,bac,ma,our9}.
In Sec.~3 we replace the Appleby-Battye (AB) scalar potential \cite{ab}
by an {\it Uplifted-Double-Well} (UDW) scalar potential. In Sec.~4 we relate 
it to $F(\car)$ supergravity. In Sec.~5 and propose a model of spontaneous 
supersymmetry (SUSY) breaking that gives rise to the UDW scalar potential. 
Sec.~6 is our conclusion.

\section{$f(R)$ Gravity and Quintessence}

The action of $f(R)$ gravity is given by
\be \lb{act}
S = \int d^4 x \, \sqrt{-g}\, f(R)
\ee
We use the natural units $\hbar=c=M_{\rm Pl}=1$, where $M_{\rm Pl}$ is the
(reduced) Planck mass, and the spacetime signature is $(+,-,-,-)$. In our
notation, a de Sitter space has a {\it negative} constant scalar curvature 
$R_0<0$.

The vacuum solutions to the theory (\ref{act}) with $R=R_0$ satisfy
\be   \lb{vacs}
R_0f'(R_0) =2 f(R_0)
\ee  

The action (\ref{act}) can be transformed to the Einstein frame, by first 
rewriting it to the form
\be \lb{act1}
S =  \int d^4 x \, \sqrt{-g}\,\left[ f'(\f)(R-\f) +f(\f) \right]~~,
\ee
where the scalar field $\f$ has been introduced. Its equation of motion reads
\be \lb{eqmf}
 f''(\f)(R-\f)=0
\ee
so that we get $\f=R$ and, hence, the original action (\ref{act}) back. 

After the conformal transformation,  
$\tilde{g}_{\m\n}= g_{\m\n} f'(\f)$, the action (\ref{act1}) reads
\be \lb{act2}
S =  \int d^4 x \, \sqrt{-g}\, \left[ -\fracmm{1}{2}\tilde{R} +
\fracmm{3}{4(f')^2}\tilde{g}^{\m\n}\pa_{\m}f'(\f)\pa_{\n}f'(\f) -
V(f'(\f)) \right]
\ee
with the scalar potential
\be \lb{spot}
V (\f) =  \fracmm{\f f'(\f) - f(\f)}{f'(\f)^2} 
\ee
The canonically normalised scalar field is given by
\be \lb{cans}
\s = - \sqrt{ \fracmm{3}{2} } \ln f'(\f)~,\quad {\rm or} \quad
f'(\f)=-\exp\left [ -\sqrt{\fracmm{2}{3} }\s \right] \equiv -e^{-y} 
\ee

Equation (\ref{spot}) is a {\it quadratic} equation w.r.t. $f'(\f)$, so that 
it can be rewritten to the form
\be \lb{diffe}
f'(R) = \fracmm{ R \pm \sqrt{R^2 -4Vf}}{2V}
\ee
where $V=V(R)$ and $f=f(R)$. Eq.~(\ref{diffe}) is the {\it inverse} problem for fixing 
the $f(R)$ function by a given scalar potential. For instance, in the case of slow-roll 
inflation, we have $V\approx const.$ so that $f'(R)\propto R$ approximately, which gives
 rise to the Starobinsky model of $(R+R^2)$ inflation \cite{star}. The present DE 
corresponds to the case $f'(R)\approx -\ha$ or $f(R)\approx -\ha R -\Lambda$ with 
$\Lambda>0$. Here we are interested in the case of the upper sign choice and 
$4Vf << R^2$ in eq.~(10). Then it reduces to $f' \approx f/R$ so that the Einstein term 
$(-\ha R)$ dominates in $f(R)$. It is enough for viability of our model in the Einstein 
approximation $|R_0|<< |R| << 1$, where $R_0$ is the present cosmic value of the scalar 
curvature.

\section{AB Function and its UDW Alternative}

The {\it trial} dark energy function proposed by Appleby and Battye in 
ref.~\cite{ab},
\be \lb{abf}
f_{\rm AB}(R) = -\ha R  +\fracmm{1}{2a}\ln \left[ \cosh (aR) +\tanh(b)\sinh(aR)
\right]~~, 
\ee
has $f(0)=0$ and is viable in the Einstein approximation. It has two positive 
parameters, $a$ and $b$, which must be fine tuned to meet observations. In the 
Newtonian limit one finds \cite{ab}
\be \lb{newt}
f_{\rm AB}(R) \approx -\ha R - \fracmm{1}{4a}\ln\left( 1+ e^{2b}\right) 
+{\cal O}\left(e^{-2a\abs{R}+2b}\right) 
\ee
so that one gets the effective cosmological constant 
$\Lambda\approx b/(2a)\approx \abs{R_0}=12H^2_0$ in the limit $b\gg 1$. To
meet observations, one should take $\Lambda^{1/4}\approx 0.0024~eV$, and 
$b\geq 30$ from imposing the local gravity constraints \cite{amens}.

The inverse function to
\be \lb{invf}
f_{\rm AB}'(R) = -\ha \left[1 - \tanh (aR+b)\right] 
\ee
is available in an {\it analytic} form, so that it is not difficult to 
calculate the effective  scalar potential from eq.~(\ref{spot}). We find
\be \lb{abpot}
V_{\rm AB}(y) = \fracmm{1}{2a}e^{2y}\left[ \ln\left(1-e^{-y}\right)
- e^{-y}\ln \left( e^y -1\right) + 2be^{-y} + C \right]~~,
\ee
where the constant $C$ is given by $C=\ln \left( e^b + e^{-b}\right) -b$. 

The parameter $a$ appears only as the overall factor in the scalar potential 
(\ref{abpot}), so we introduce $v_{\rm AB} =V_{\rm AB}/V_0$ with 
$V_0=(2a)^{-1}$. The function $v_{\rm AB}(y)$ has only one parameter $b$, and 
its profile (at $b=1.5$) is shown in Fig.~1.

\begin{figure}
\includegraphics[height=3.2in,width=5.3in]{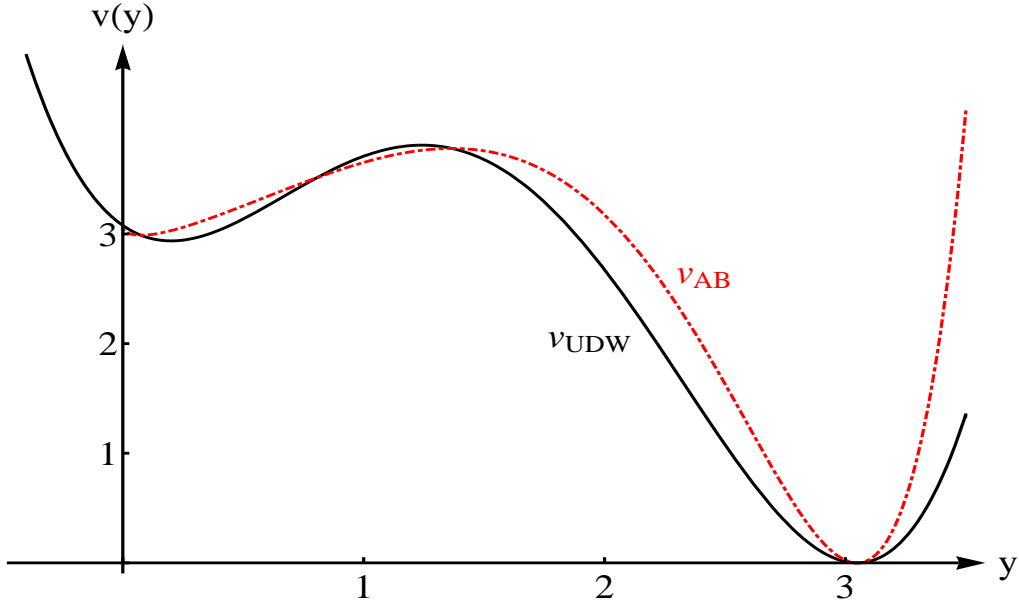}
\caption{\small The Uplifted-Double-Well scalar potential versus the 
Appleby-Battye scalar potential $(b=1.5)$.}
\label{fig1}
\end{figure}

Away from $y=0$ on its right-hand-side (when $b>1$ ) the function $v_{\rm AB}(y)$ has 
two minima and one maximum, so  it can be well approximated by an 
{\it Uplifted-Double-Well} (UDW) scalar potential of the Higgs-type (see Fig.~1),
\be \lb{uwdp} v_{\rm UDW}(y)=\fracmm{1}{4}\left[ (y-y_0)^2-v^2\right]^2
+\fracmm{\m^2}{2}\left[ (y-y_0)-v\right]^2 ~~,
\ee
where we have introduced three real parameters $(y_0,v,\m)$. The extrema of a
quartic scalar potential are given by roots of a cubic equation. In the case
(\ref{uwdp}) one root $(y_3)$ is given by $y_c=y_0+v$, whereas the remaining
two roots $y_{\pm}$ are the roots of a quadratic equation. We find 
\be \lb{qroots}
y_{\pm}-y_0= \ha \left[ -v \pm \sqrt{v^2-4\m^2} \right] 
\ee

By demanding the local minima of  $v_{\rm UDW}$ to coincide with those of 
$v_{\rm AB}$ we find
\be \lb{fixpa}
\m^2=b^{-1}-\ha b^{-4}~~,\quad y_0=b-\ha b^{-2}~~,\quad v=b+\ha b^{-2}
\ee
For large $b\gg 1$ the potential barrier between the two vacua (de Sitter and
Minkowski) {\it exponentially grows} as $e^{2(b-1)}$, while the constant $C$ 
goes to zero. 

The UDW scalar potential (\ref{uwdp}) is analytic, is bounded from below, and is
non-negative. It has the absolute minimum corresponding to the flat (Minkowski) vacuum, 
and another minimum corresponding to the de Sitter vacuum that can be identified with 
an accelerating universe. Those vacua are separated by the high potential  barrier, so 
that the lifetime of the universe in the {\it meta-stable} de Sitter vacuum can be 
larger than its age.

The AB scalar potential (\ref{abpot}) has a non-analytic behaviour at $y=0$ which 
corresponds to the infinite scalar curvature $R$. The UDW scalar potential (\ref{uwdp}) 
is regular at $y=0$ (and also for $y<0$), while $y=0$ corresponds to the low space-time 
curvature of the order $R_0$. Hence, the AB and UDW functions are drastically different 
in the regime of the high space-time curvature where both models cannot be trusted. 
However, they are almost the same in the Einstein regime $|R_0|<< |R| << 1$, which is 
enough for our purpose.

\section{$F(\car)$ Supergravity and Quintessence}

The action of $F({\cal R})$ supergravity in the chiral (curved) ${\cal N}=1$ 
superspace of $(1+3)$-dimensional spacetime, which was proposed in 
ref.~\cite{Gates}, reads
\be
S=\int d^4 x\,d^2 \theta\, {\cal E}\, F(\car)+{\rm H.c.}\,,
\label{frsup}
\ee
where $F(\car)$ is a holomorphic function of the covariantly chiral scalar 
curvature superfield ${\cal R}$, and ${\cal E}$ is the chiral superspace 
density. The scalar curvature $R$ appears as the field coefficient at the 
$\theta^2$ term in the superfield ${\cal R}$.~\footnote{See Ref.~\cite{Kreview}
 for details about our notation and $F({\cal R})$ supergravity.} The action
(\ref{frsup}) is equivalent to 
\be 
S=\int d^4 x\,d^2 \theta\, {\cal E}\, \left[ -\cy \car +Z(\cy) \right] 
+{\rm H.c.}\,,
\label{frsup2}
\ee
where we have introduced the new covariantly chiral scalar superfield $\cy$ 
and the new holomorphic function $Z(\cy)$ related to the function $F$ as 
\be \label{zfun}
F(\car) = -\car \cy(\car) + Z(\cy(\car))\,.
\ee
The equation of motion of the superfield $\cy$, which follows from the 
variation of the action (\ref{frsup2}) with respect to $\cy$, has the algebraic
 form
\be \label{yem}
\car = Z'(\cy) 
\ee
so that the function $\cy(\car)$ is obtained by inverting the function $Z'$.  
Substituting the solution $\cy(\car)$ back into the action (\ref{frsup2}) 
yields the original action (\ref{frsup}) because of Eq.~(\ref{zfun}). We also
have
\be \label{leg}
\cy =- F'(\car)
\ee
The inverse function $\car(\cy)$ always exist under the physical condition  
$F'(\car) \neq 0$. 

The kinetic terms of $\cy$ are obtained by using the (Siegel) identity 
\be \label{siegel}  
\int d^4 x\,d^2 \theta\, {\cal E}\, \cy \car  +{\rm H.c.}=
\int d^4 x \, d^4 \theta \, E^{-1} (\cy +\bar{\cy} ) \, ,
\ee
where $E^{-1}$ is the full curved superspace density. Therefore, the K\"ahler 
potential reads 
\be \label{kahl}
K = -3 \ln \left( \cy + \bar{\cy} \right) 
\ee
and gives rise to the kinetic terms
\be \label{kinc}
{\cal L}_{\rm kin} =  \left.
\fracmm{\pa^2 K}{\pa\cy \pa\bar{\cy}}\right|_{\cy=Y}
\pa_{\mu} Y\pa^{\mu}\bar{Y} = 3 \fracmm{ \pa_{\mu}Y\pa^{\mu}\bar{Y} } 
{(Y +\bar{Y})^2}
\ee
The kinetic terms (\ref{kinc}) represent the {\it non-linear sigma model} 
\cite{ketovbook}  with the hyperbolic target space of (real) dimension two, 
whose metric is known as the standard (Poincar\'e) metric with $SL(2,{\bf R})$ 
isometry.

In the decoupling limit of supergravity, the effective scalar potential 
$V(Y,\bar{Y})$ of a complex scalaron $Y$ is derived from eq.~(\ref{frsup2}) 
when keeping only scalars (i.e. ignoring their spacetime derivatives
together with all fermionic contributions) and eliminating the auxiliary 
fields near the minimum of the scalar potential. One finds \cite{shinji} 
\be \label{scals}
V = \fracmm{21}{2} \left|  Z'(Y) \right|{}^2 =  \fracmm{21}{2} \left|  
\car(Y) \right|{}^2 
\ee
that gives rise to the chiral superpotential
\be \label {chipot} 
W (\cy) = \sqrt{\fracmm{21}{2}} Z(\cy)\,.
\ee
The superfield equations (\ref{kahl}) and (\ref{chipot}) are model-independent,
i.e. they apply to any function $F({\cal R})$ in the large $M_{\rm Pl}$ limit,
near the minimum of the scalar potential with the vanishing cosmological 
constant. After the holomorphic superfield redefinition 
\be \lb{srdef}
\cy=\exp\left( \sqrt{\fracmm{2}{3}}\F\right)
\ee 
the kinetic terms (\ref{kinc}) begin with the canonical term $\bar{\F}\F$ or 
$\pa_{\m}\bar{\f}\pa^{\m}\f$ where $\left.\F\right|=\f$. Accordingly, the 
chiral superpotential of $\F$ is given by $W(\F)=W(\cy(\F))$.

\section{Spontaneous SUSY breaking model} 

The DE in the present universe, like any other (positive) non-vanishing vacuum 
energy, requires a spontaneous SUSY breaking. The latter can occur in the 
hidden matter sector (beyond the MSSM), which must include the scalaron 
superfield $\F$ in our scenario. Since scalaron has the universal interaction
with the gravitational strength to all matter fields, it is also the natural 
messenger of the {\it gravitational mediation} of the SUSY breaking from the
hidden sector to the visible (matter) sector.~\footnote{The idea of the 
gravitational mediation of SUSY breaking was proposed in ref.~\cite{can}.}

The simplest (Wess-Zumino-type) model of the hidden sector, leading to 
spontaneous SUSY breaking,
consists of three chiral superfields, $\F_1$, $\F_2$ and $\F_3$.
Let us choose the chiral superpotential of $\F_1$ in the form
\be \lb{chsp1}  
W_1(\F_1) = \l^{1/2}\left( \fracmm{1}{6}\F^3_1 - \fracmm{1}{2}v^2\F_1\right)~, 
\ee
with two real (positive) parameters $\l$ and $v$. It gives rise to the scalar
potential 
\be \lb{sp1}
V_1(\f_1) = \left.\abs{\fracmm{\pa W_1}{\pa\F_1}}^2\right| 
=\fracmm{\l}{4}\abs{\f^2_1-v^2}^2
\ee
where $\left.\F_1\right|=\f_1$. Similarly, the chiral superpotential of $\F_2$ 
in the form
\be \lb{chsp2}  
W_2(\F_2) = \fracmm{\m}{\sqrt{2}}\left( \fracmm{1}{2}\F^2_2 - u \F_2\right)~, 
\ee
with two real (positive) parameters $\m$ and $u$, gives rise to the scalar
potential 
\be \lb{sp2}
V_2(\f_2) =\left. \abs{\fracmm{\pa W_2}{\pa\F_2}}^2\right| =
\fracmm{\m^2}{2}\abs{\f_1-u}^2
\ee
Therefore, the use of the chiral superpotential
\be \lb{ospot}
 W(\F_1,\F_2,\F_3)=W_1(\F_1) + W_2(\F_2) + \F_3(\F_2-\F_1)
\ee
gives rise to the scalar potential
\be \lb{raif}
V(\f) = V_1(\f) + V_2(\f) = \fracmm{\l}{4}\abs{\f^2-v^2}^2 +
\fracmm{\m^2}{2}\abs{\f-u}^2
\ee
where $\f=\left.\F\right|$ and $\F=\F_1=\F_2$. The scalar potential 
(\ref{raif}) is the complex extension of the UDW scalar potential (\ref{uwdp}),
where ${\rm Re}(\f)=y-y_0$ and $u=v$.

The superpotential (\ref{ospot}) is the  particular example of the
O'Raifeartaigh-type models of spontaneous SUSY breaking \cite{rai}. Indeed,
the system of equations 
\be \lb{sysn}
\fracmm{\pa W_1}{\pa\F_1}=\fracmm{\pa W_2}{\pa\F_2}=\fracmm{\pa W_3}{\pa\F_3}
=0
\ee 
does not have a solution when $u\neq v$, which gives rise to a positive vacuum
energy. When $u=v$, we have a stable (Minkowski) vacuum with unbroken SUSY and 
the vanishing vacuum energy, but also a metastable (de Sitter) vacuum with DE.

\section{Conclusion}

Our main results are given by eqs.~(\ref{diffe}), (\ref{abpot}), (\ref{uwdp}), 
(\ref{fixpa}) and (\ref{ospot}). We replace the effective scalar potential (\ref{abpot})
 associated with the {\it ad hoc} AB function (\ref{abf}) by the Higgs-type scalar 
potential (\ref{uwdp}) that gives rise to a meta-stable accelerating universe. We 
propose the specific (O'Raifertaigh-type) model of the hidden sector leading to 
spontaneous SUSY breaking and the UDW scalar potential, in terms of three chiral scalar 
superfields with the chiral superpotential (\ref{ospot}). In our appproach the chiral 
scalaron superfield is the universal messenger of the gravitational mediation of SUSY 
breaking to the visible sector (Standard Model) of elementary particles.

\section*{Acknowledgements}

SVK is supported by the World Premier International Research Center Initiative 
(WPI Initiative), MEXT, Japan. SVK is grateful to A.~D. Dolgov and T.~T. 
Yanagida for discussions.

\end{document}